\input{aipcheck}

\documentclass[
    ,final            % use final for the camera ready runs
  ]
  {aipproc}

\layoutstyle{6x9}
\usepackage{lineno} % specify the line number in the margin

\newcommand {\pT}{\ensuremath{p_{\mathrm{T}}} }

\newcommand {\dEdx}{\mathrm{d}E/\mathrm{d}x }
\newcommand {\PbPb}{\mbox{Pb--Pb} }
\newcommand {\AuAu}{\mbox{Au--Au} }

\newcommand{\sqrtsNN} {\mbox{$\sqrt{s_{\mathrm{NN}}}$}}

\newcommand {\mass} {\mbox{\rm MeV$\kern-0.15em /\kern-0.12em c^2$}}
\newcommand {\tev} {\mbox{${\rm TeV}$}}
\newcommand {\gev} {\mbox{${\rm GeV}$}}

\newcommand {\mom} {\mbox{\rm GeV$\kern-0.15em /\kern-0.12em c$}}
\newcommand {\gmom} {\mbox{\rm GeV$\kern-0.15em /\kern-0.12em c$}}
\newcommand {\mmass} {\mbox{\rm MeV$\kern-0.15em /\kern-0.12em c^2$}}
\newcommand {\gmass} {\mbox{\rm GeV$\kern-0.15em /\kern-0.12em c^2$}}
\newcommand {\mmom} {\mbox{\rm MeV$\kern-0.15em /\kern-0.12em c$}}

\newcommand{\piplus}{\mbox{$\mathrm {\pi^{+}}$}}
\newcommand{\piminus}{\mbox{$\mathrm {\pi^{-}}$}}
\newcommand{\pich}{\mbox{$\mathrm {\pi^{\pm}}$}}

\newcommand{\pbar} {\mbox{$\mathrm {\overline{p}}$}}
\newcommand{\K}{\mbox{$\mathrm {K}$}}
\newcommand{\Kzs}{\mbox{$\mathrm {K^0_S}$}}
\newcommand{\Kplus}{\mbox{$\mathrm {K^{+}}$}}
\newcommand{\Kminus}{\mbox{$\mathrm {K^{-}}$}}
\newcommand{\Kch}{\mbox{$\mathrm {K^{\pm}}$}}
\newcommand{\rmLambda}{\mbox{$\mathrm {\Lambda}$}}
\newcommand{\rmAlambda}{\mbox{$\mathrm {\overline{\Lambda}}$}}

\newcommand{\Jpsi} {\mbox{J\kern-0.05em /\kern-0.05em$\psi$}}

\newcommand{\pp} {\mbox{$\mathrm {pp}$}}
\newcommand{\rmBeta}{\mbox{$\mathrm {\beta}$}}
\newcommand{\rmTfo}{$T_{\mathrm{fo}}$}

\begin{document}
%%\setpagewiselinenumbers
%%\modulolinenumbers[5]
%%\linenumbers

\title{Strange and identified hadron production at the LHC with ALICE }


\author{L. S. Barnby for the ALICE Collaboration}{
  address={School of Physics and Astronomy, University of Birmingham, Birmingham. B15 2TT. UK}
}

%\author{<author3>}{
%  address={<common address for author2 and author3>}
 % ,altaddress={<author1 address>} % additional visiting address
%}

\begin{abstract}
 The ALICE detector was designed to identify hadrons over a wide range of transverse momentum at mid-rapidity. Here measurements of light charged ($\pi$, \K, p) and neutral (\rmLambda, \Kzs) hadrons in \PbPb collisions at \sqrtsNN~= 2.76 \tev~are presented with additional data from a pp reference at $\sqrt{s} = 7$ \tev. Such measurements are crucial for understanding the properties of the fireball produced in heavy-ion collisions at the LHC. The particle-type dependence of the spectra and the yields of particles extracted give information on the expansion dynamics and chemical composition respectively. In addition studying the ratio of baryons to mesons may help in understanding the mechanisms by which hadronisation takes place. We find that, when comparing to data from \sqrtsNN~= 200 \gev~Au+Au collisions at RHIC, a more strongly expanding system is created with a similar relative population of hadron species. We also see that collective effects or complex mechanisms responsible for a relative enhancement of baryons have an influence at a much higher \pT than was previously seen.
 \end{abstract}

\maketitle

%%%%%%%%%%%%%%%%%%%%%%%%%%%%%%%%%%%%%%%%%%%%
%% MAINMATTER
%%%%%%%%%%%%%%%%%%%%%%%%%%%%%%%%%%%%%%%%%%%%

\section{Introduction}

  The aim of relativistic heavy-ion collision experiments is to detect and understand the properties of the bulk QCD matter created in these collisions. Past experiments have shown that the ability to perform measurements differentially with respect to the identity of the final state hadrons is crucial to a full understanding of the evolution and dynamics of the produced fireball \cite{BraunMunzinger199543,BraunMunzinger19961,KolbRapp}. Such measurements have also revealed anomalies challenging the detailed modelling of the collision \cite{Adler:2003fk,Adams:2006wk}. The ALICE experiment was designed with the goal of maximising the particle identification capability using transition and Cherenkov radiation detectors, calorimetry and, in particular for the analysis presented here, identifying the most abundant species of charged hadrons over a wide range of \pT at mid-rapidity using $\dEdx$ and time-of-flight techniques \cite{1748-0221-3-08-S08002}. The excellent tracking down to low \pT also allows the reconstruction of weakly decaying neutral strange particles via their charged decay modes. 
  
\section{Experiment}

The ALICE central barrel performs tracking of charged particles in a 0.5 T magnetic field using a Time Projection Chamber (TPC) and Inner Tracking System (ITS). Particles with large enough \pT pass through the outer wall of the TPC and can go on to hit a surrounding Time-of-Flight detector (TOF).
\PbPb events were collected using a minimum bias trigger and several million events are used in this analysis. The \PbPb data sample can be separated into centrality bins using the event-wise multiplicity in the VZERO forward scintillator detectors in combination with a Monte Carlo Glauber study \cite{Aamodt:2011qy}. Charged particle identification is achieved using two techniques. The specific energy loss, $\dEdx$, can be calculated for each track from the ionisation in the TPC gas (or ITS silicon) and compared to theoretical values from the Bethe-Bloch formula which predicts the regions in momentum where $\pi$, \K, and p signals can be separated. This separation between species can be used at low \pT~but near to the minimum of $\dEdx$  all three species are merged. In this range however the TOF can separate these species so a combined \pT~spectrum can be extracted \cite{Aamodt:2010uq}. In this analysis the primary yield of charged particles is reported; that is those emerging directly from the collision or the decay of short-lived resonances and not the charged particles from the weak decay of strange hadrons nor secondaries from the material. These are both excluded using the distribution of the distance of closest approach to the primary interaction vertex, which can be fitted to a template obtained from Monte Carlo events, where the origin of the particle is known. The decay of neutral strange particles decaying into charged daughters; $\Lambda\rightarrow p \piminus$ and $\Kzs \rightarrow \piplus \piminus$, can be reconstructed and the invariant mass distributions used for identification. The analysis follows the method used for \pp~collisions but tighter cuts are made to further reduce the combinatorial background in \PbPb events \cite{Aamodt2011Strange}. For both neutral and charged particle analyses the spectra are corrected using efficiencies from Monte Carlo events having equivalent mean multiplicities. 

%\paragraph{<A subsubsubsection>}

\section{Results}
\subsection{Charged Particles}
\begin{figure}
   \begin{minipage}{\columnwidth}
\centering
  \includegraphics[height=.27\textheight]{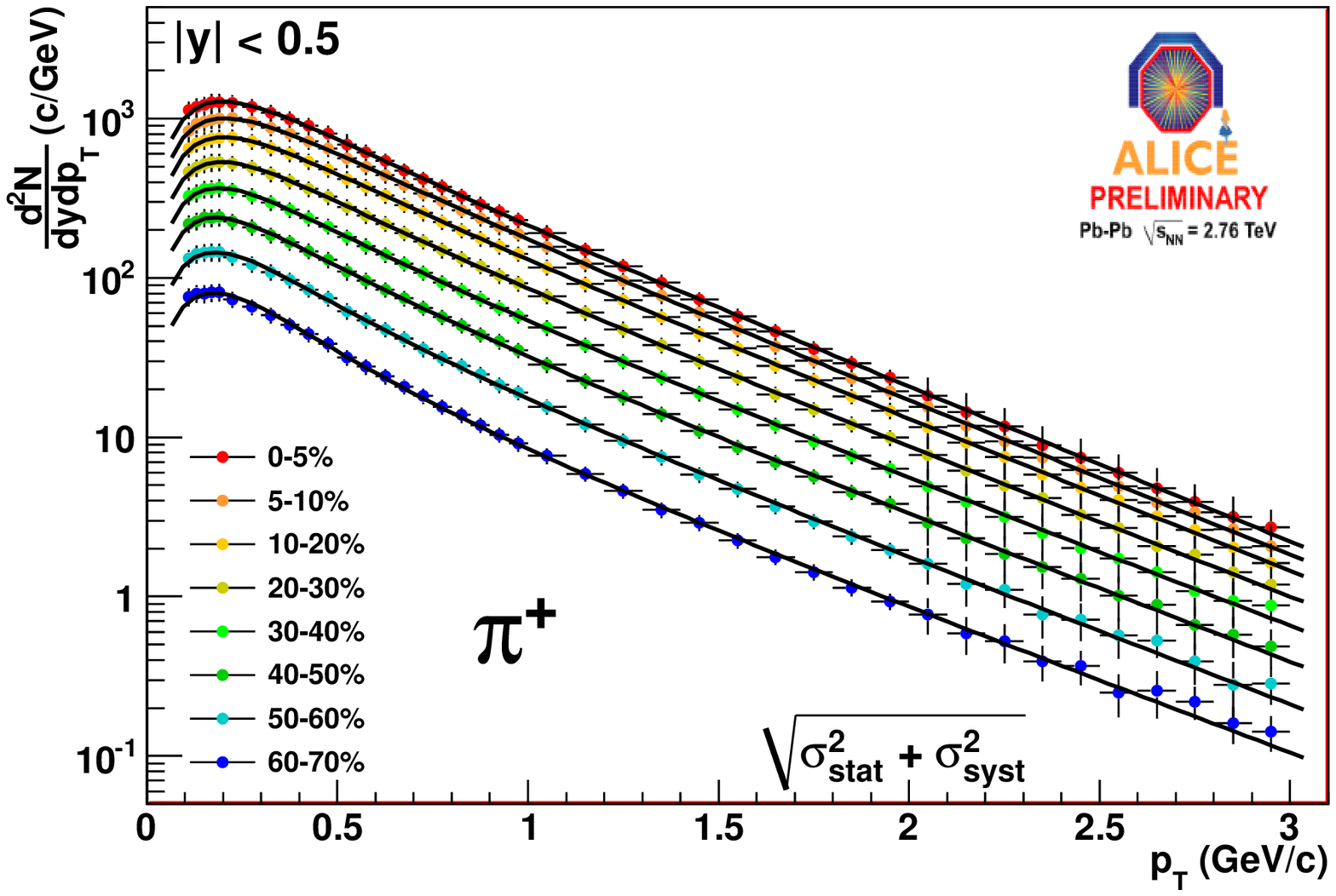}
  \caption{The centrality-selected \pT spectra for identified \piplus (top) \Kplus (middle) and p (bottom). Fits are to a parameterised blast wave.}
%\end{figure}
%\begin{figure}
  \includegraphics[height=.27\textheight]{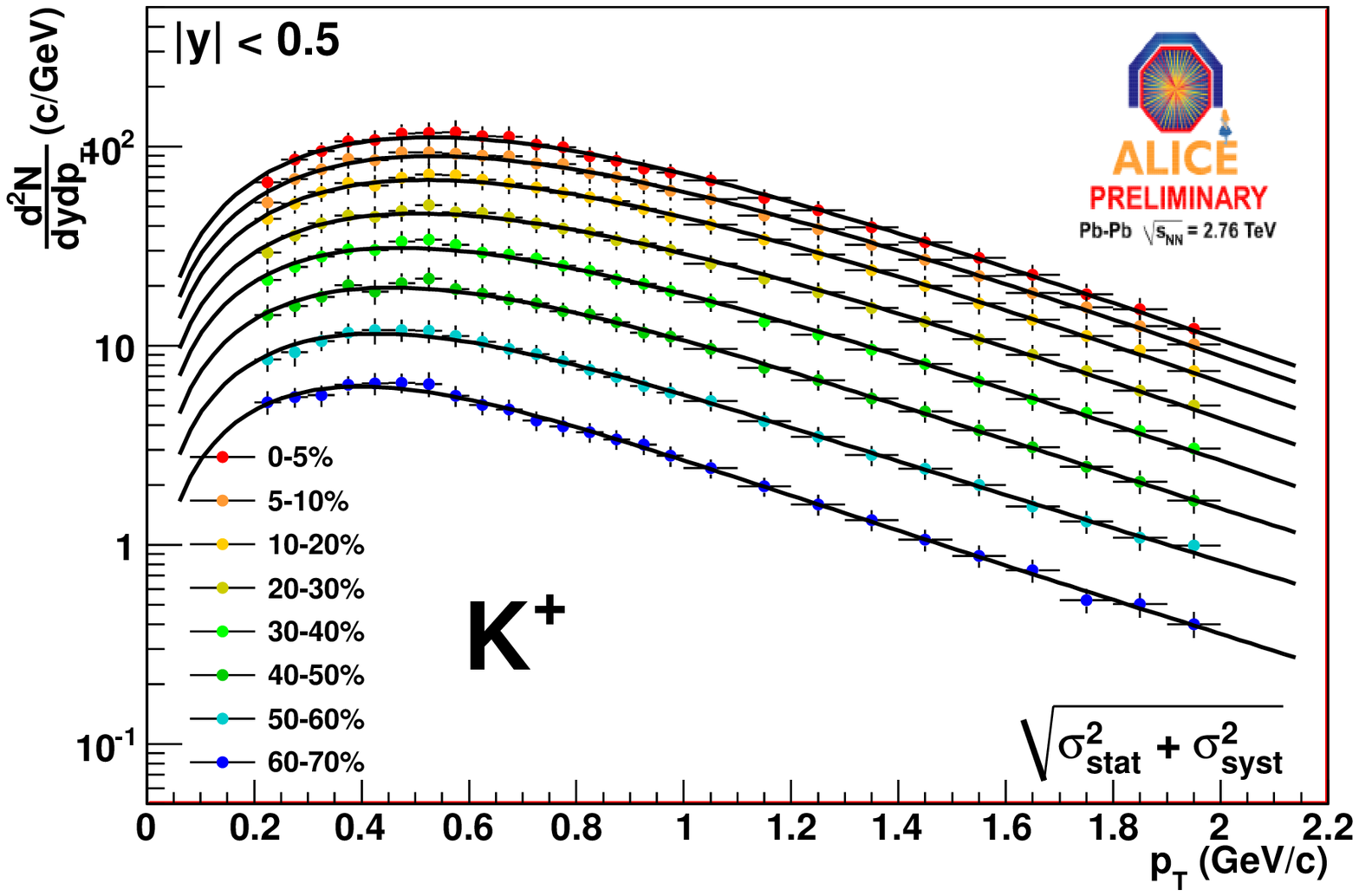}
 % \caption{Picture to fixed height}
%\end{figure}
%\begin{figure}

  \includegraphics[height=.27\textheight]{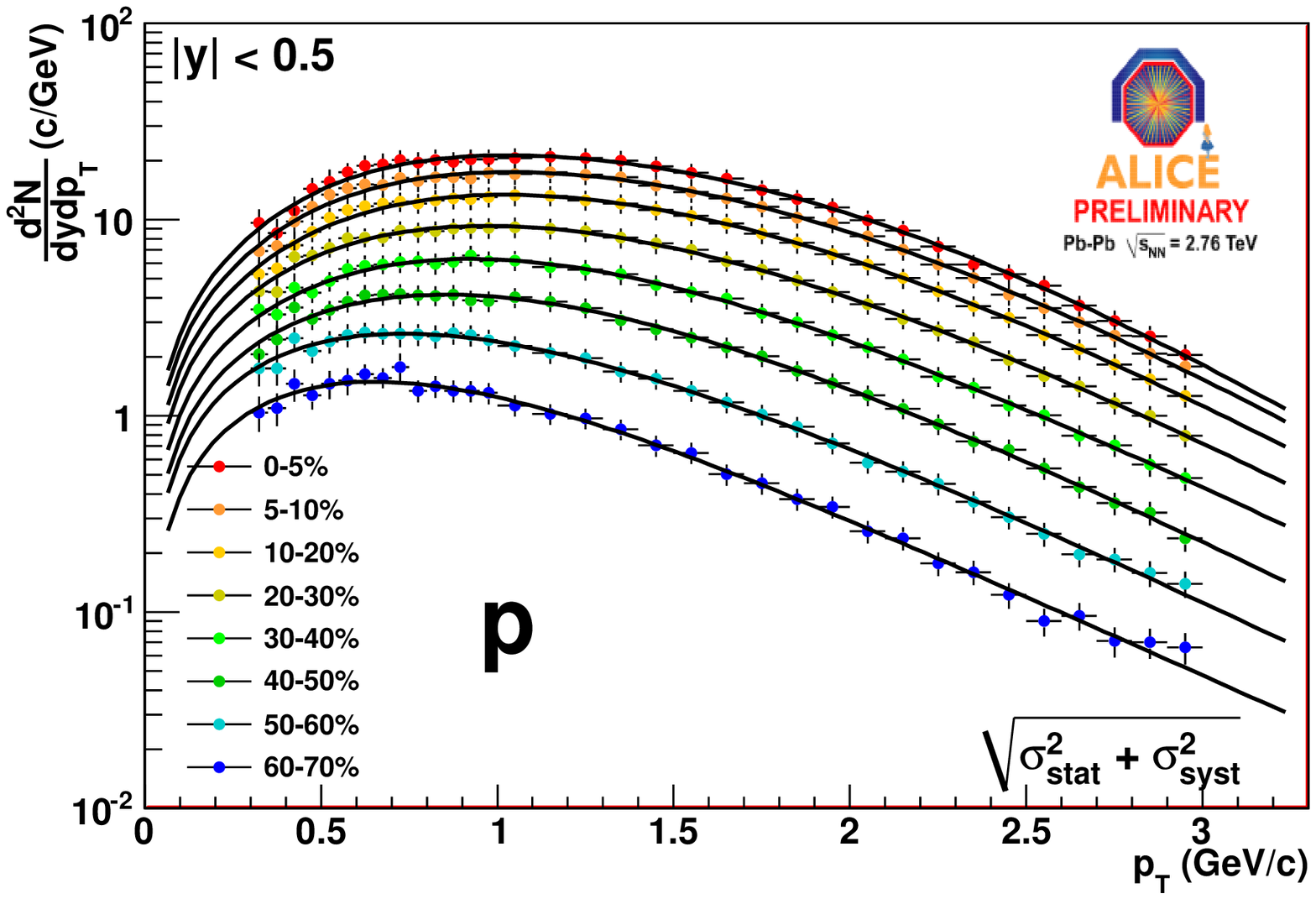}
 % \caption{Picture to fixed height}
\label{thespectra}
\end{minipage}
\end{figure}

\begin{figure}
\includegraphics[height=.30\textheight]{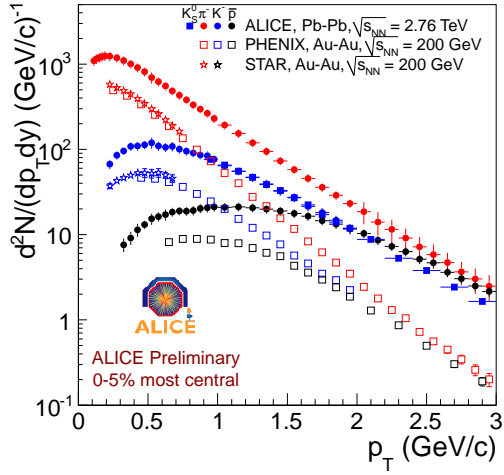}
\caption{The \pT spectra for  \piminus, \Kzs, \Kminus, and \pbar~for the most central \PbPb (0-5\%) collisions (solid markers) plotted with those measured in \sqrtsNN~= 200 \gev \AuAu collisions (open symbols.)}
\label{spectraRHIC}
\end{figure}

\begin{figure}
\includegraphics[height=.28\textheight]{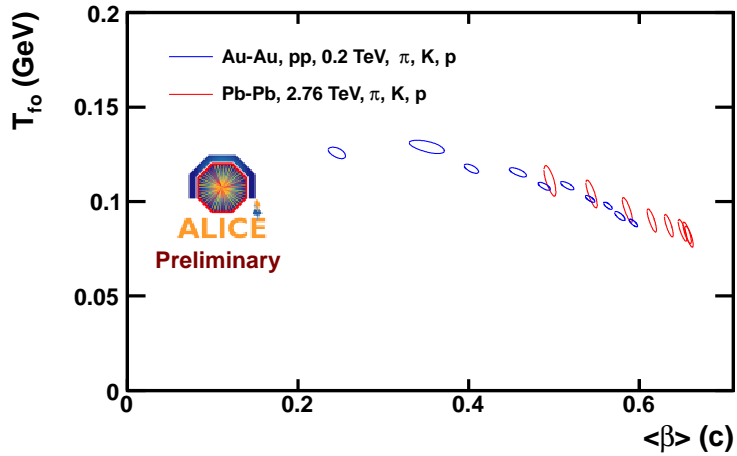}
\caption{1-$\sigma$ contours in the T-\rmBeta~plane from a simultaneous fit of a parameterised blast wave function to the \pich, \K, and p \pT spectra  for various centrality classes.  \PbPb collisions from the ALICE experiment in red, \AuAu collisions from the STAR experiment in blue. Most central data lie to the right.}
\label{bwave}
\end{figure}

The combined \pT~spectra are obtained for each of eight centrality bins for \pich, \Kch, p and \pbar~and are shown, for positive particles only, in figure \ref{thespectra}. The most noticeable features are: the dramatic change in the shape of the spectra going from $\pi$ through \K~to p; and the shifting of the most probable values to higher $\pT$, particularly for p but also for \K. A direct comparison of the most central spectra to \AuAu data at \sqrtsNN~= 200 \gev~is made in figure \ref{spectraRHIC}. This shows how the spectra at LHC energy are much less steeply falling.  A first attempt at quantifying the changes using a parameterised blast wave function \cite{Schnedermann:1993uq} was made. The resulting fit parameters for the freeze-out temperature, \rmTfo, and mean transverse velocity, \rmBeta,  are shown in figure \ref{bwave} as 1-$\sigma$ contours for each centrality class. Fits ranges 0.3--1.0 \gmom,  0.2--1.5 \gmom~and 0.3--3.0 \gmom~were used for $\pi$, \K~p respectively in order to avoid the region where a hard component of the spectum might be expected and, at low $\pT$, to avoid a strong contribution of resonances to $\pi$. There appears to be a larger \rmBeta, corresponding to stronger flow, than observed by STAR at lower energy \cite{Adams2005102}. \rmTfo~is very sensitive to the fit range so any change with respect to RHIC needs further study. A blast wave was also fitted to each individual spectrum in order to obtain $\pT$-integrated yields, including the unmeasured part. These can be used to form the ratios p/$\pi$ and $\K/\pi$ for each centrality bin. The ratio  p/$\pi$ is almost constant with centrality and is consistent with similar measurements in \AuAu collisions at RHIC \cite{Adler:2004fj}. The ratio $\K/\pi$ shows a small rise from \pp~and peripheral collisions to the most central collisions and is also consistent with previous lower energy data \cite{Abelev:2009kx}.

 %%%%%%%%%%%%%%%%%%%%%%%%%%%%%%%%%%%%%%%%%%%%
%% Sample figure:
%%
%% The option [height=...] scales the picture to the given height,
%% without it it would be printed at its nominal size
%%%%%%%%%%%%%%%%%%%%%%%%%%%%%%%%%%%%%%%%%%%%

\subsection{Neutral Particles}
\begin{figure}
   \begin{minipage}{\columnwidth}
\centering
\includegraphics[height=.29\textheight]{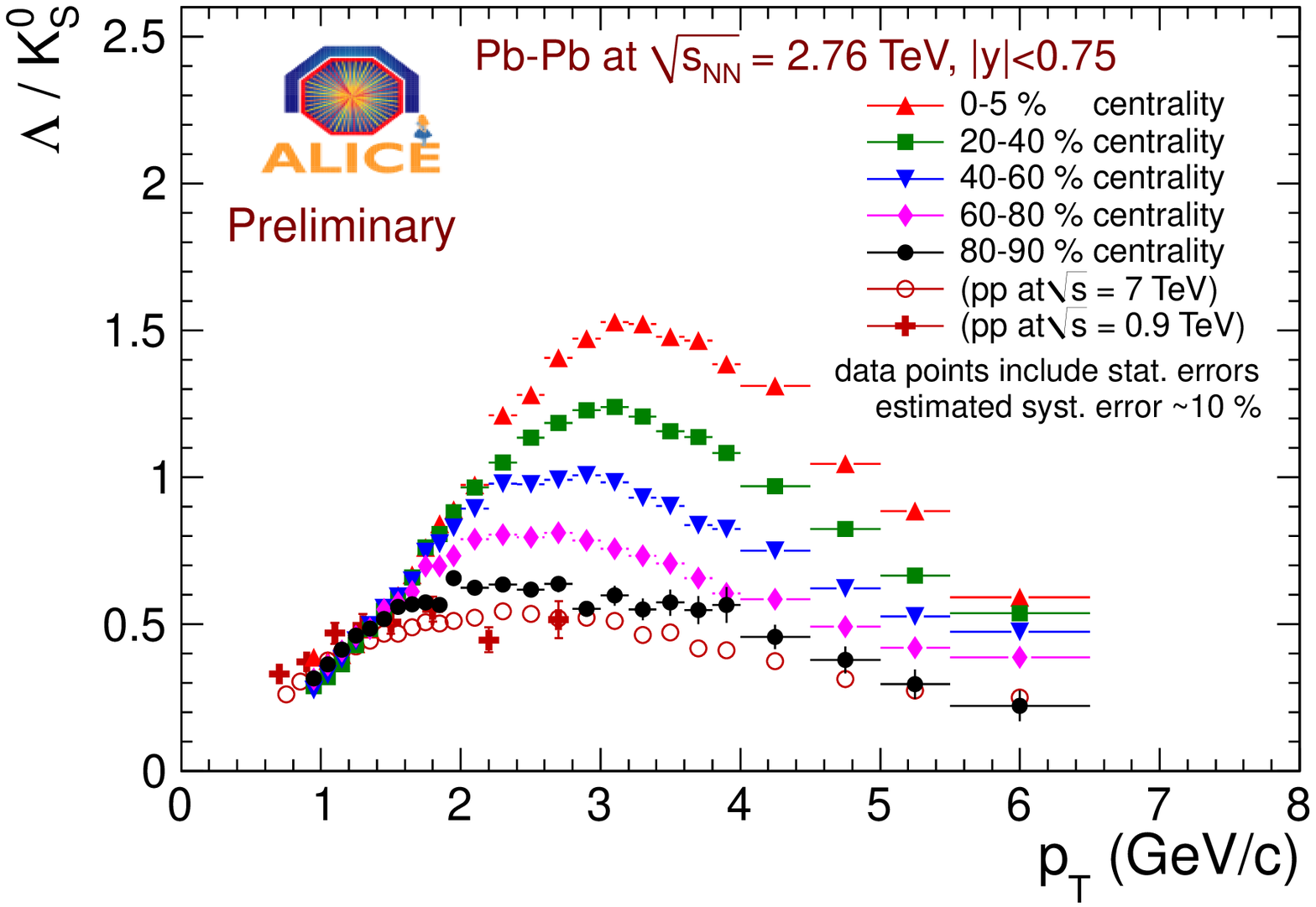}
\includegraphics[height=.273\textheight]{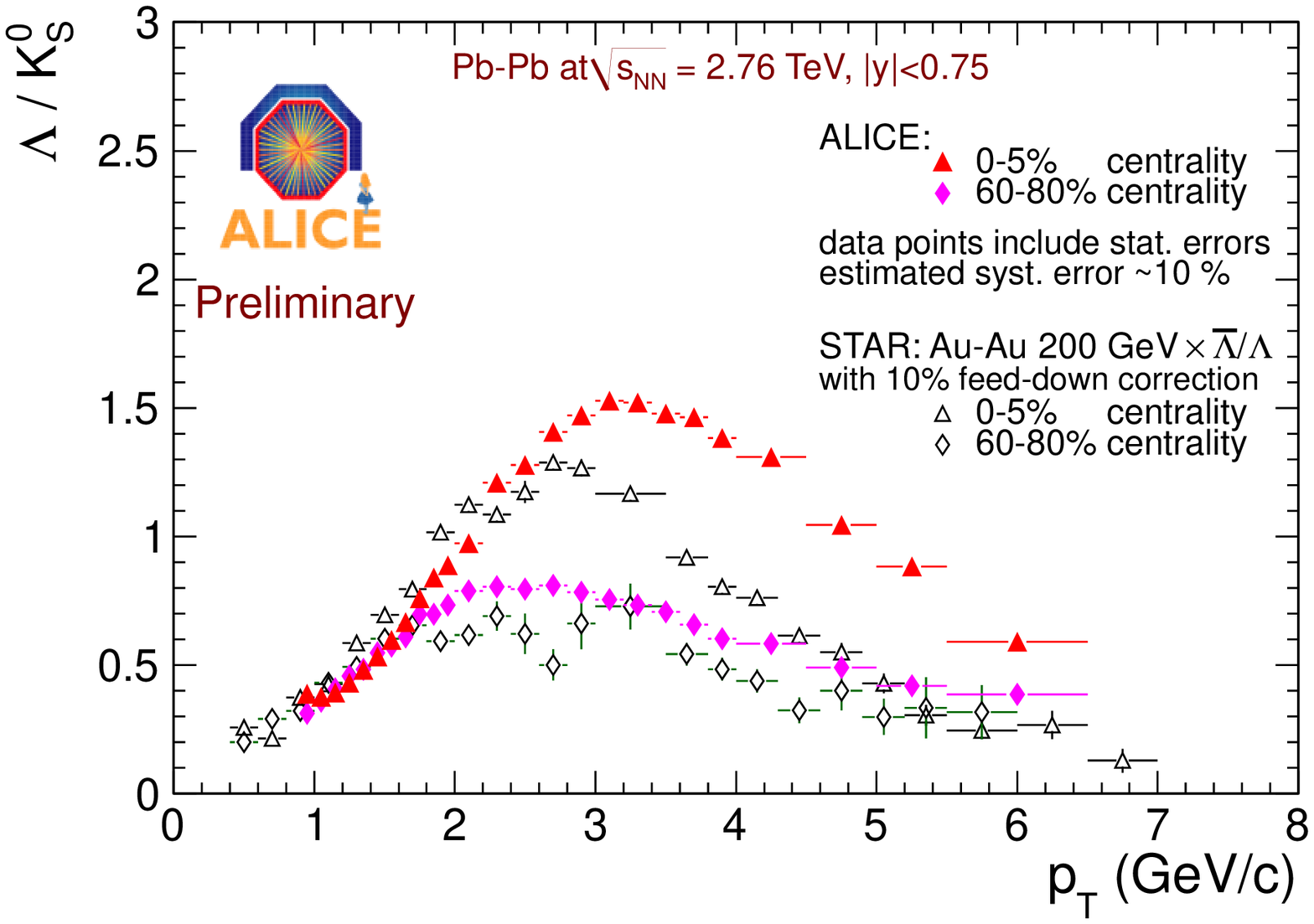}
\caption{(Upper panel.) The ratio of \rmLambda~to \Kzs~as a function of \pT for five centrality classes in \PbPb collisions. Also shown the same ratio in \pp~collisions at two energies. (Lower panel.) A comparison between the ratio measured by ALICE (solid markers) and STAR (open symbols) for selected centralities.}
\label{LKRatio}
\end{minipage}
\end{figure}

The \pT spectra of \rmLambda~and \Kzs~have also been extracted for each centrality bin. As the systematic uncertainties on the efficiency correction are still under study the preliminary spectra themselves are not yet ready. However the study reveals that the ratio of the efficiencies for each particle, as a function of \pT, is rather stable with respect to changing the centrality of the collision. In particular in the \pT range 2.5-5.5 \gmom~the variation of the efficiency ratio between the most central and the most peripheral centrality selections is below 2\%. This allows the \rmLambda / \Kzs~ratio  to be calculated with an estimated systematic uncertainty of 10\% and the resulting curves for each centrality are shown in figure \ref{LKRatio} (upper). Also shown are the ratios in \pp~collisions at $\sqrt{s} = 0.9$ and 7 \tev~\cite{Aamodt2011Strange}. The \pp~data demonstrate that in the \tev~range the maximum value of the ratio is almost constant and it is reasonable to assume that \pp~collisions at $\sqrt{s} = 2.76$ \tev~would show the same maximum. Taking this \pp~baseline the ratio is observed to have a maximum which rises strongly going to peripheral and then to central events, with a total increase up to a factor of three. The value of \pT at which the maximum is reached is also increasing by several hundred \mmom. The data are compared to a similar measurement previously made by STAR in figure \ref{LKRatio} (lower) \cite{0954-3899-32-12-S13}. To facilitate the comparison the lower energy data were scaled by the \rmAlambda/\rmLambda~ratio measured for each centrality \cite{Adams:2007lr}, assuming it is constant in $p_{\mathrm{T}}$, because it has previously been noted that there is a $\sqrt{s}$-dependence of the ratio \cite{Aggarwal:2011lr}, presumably due to the change in the baryo-chemical potential. The ALICE data were not scaled in this way because the anti-baryon/baryon ratio in LHC collisions is very close to one \cite{Aamodt:2010yq}. The ratio in peripheral 60-80\% collisions is very similar in shape for the two collision systems with only a small change in the magnitude. In the most central 0-10\% however the shape is quite different with the enhancement of the \rmLambda~extending to a much larger \pT in the higher energy data. This is qualitatively in agreement with some predictions \cite{springerlink:10.1140/epjcd/s2004-04-026-6}.

\section{Conclusions}

\PbPb collisions at \sqrtsNN~= 2.76 \tev~reveal a number of similarities to \AuAu collisions at RHIC;  the ratios of the yields are the same within the experimental uncertainties, the spectra are compatible with a strong collective motion which increases going to more central collisions and there is a growth of the \rmLambda / \Kzs~ratio in the  \pT region 2-4 \gmom, also with centrality. There are however some notable differences; the \pT spectra are much flatter giving a transverse flow velocity in a blast wave parameterisation  10\% larger than that in \sqrtsNN~= 200 \gev \AuAu collisions and the enhanced baryon-to-meson ratio extends to a \pT of around 6 \gmom. This may imply that the influence of particles participating in the collective dynamics of the system extends to a higher \pT than has previously been observed.

%%%%%%%%%%%%%%%%%%%%%%%%%%%%%%%%%%%%%%%%%%%%%%%%
%% BACKMATTER
%%%%%%%%%%%%%%%%%%%%%%%%%%%%%%%%%%%%%%%%%%%%%%%%

%\begin{theacknowledgments}
%  Infandum, regina, iubes renovare dolorem, Troianas ut opes et
%  lamentabile regnum cruerint Danai; quaeque ipse miserrima vidi, et
%  quorum pars magna fui. Quis talia fando Myrmidonum Dolopumve aut duri
%  miles Ulixi temperet a lacrimis?
%\end{theacknowledgments}



\bibliographystyle{aipproc}   % if natbib is available

\bibliography{EPICBarnby}

\IfFileExists{\jobname.bbl}{}
 {\typeout{}
  \typeout{******************************************}
  \typeout{** Please run "bibtex \jobname" to optain}
  \typeout{** the bibliography and then re-run LaTeX}
  \typeout{** twice to fix the references!}
  \typeout{******************************************}
  \typeout{}
 }

\end{document}